\documentclass[12pt]{article}
\usepackage{graphics}
\usepackage{amssymb}

\newcommand{\R}{\mathbb{R}}
\newcommand{\Z}{\mathbb{Z}}

\newcommand{\DD}{\mathcal{D}}
\newcommand{\PP}{\mathcal{P}}
\newtheorem{claim}{Claim}[section]
\newtheorem{theorem}[claim]{Theorem}
\newtheorem{conjecture}[claim]{Conjecture}
\newtheorem{proposition}[claim]{Proposition}
\newenvironment{proof}[1][Proof]{\vspace{-.8ex}\textsl{#1:}$\:$\small }
      {\ \hspace{-2pt}\rule[-.5pt]{3.7pt}{7.5pt}\vspace{.8ex}}

\begin{document}

\title{An isoperimetric problem for point interactions}
\author{Pavel Exner}
\date{}
\maketitle
\begin{center}
{\small \em
 Department of Theoretical Physics, Nuclear
Physics Institute, \\ Academy of Sciences, 25068 \v{R}e\v{z} near
Prague, Czechia, and \\ Doppler Institute, Czech Technical
University, B\v{r}ehov{\'a} 7, \\
11519 Prague, Czechia \\ \rm exner@ujf.cas.cz}
\end{center}

\vspace{8mm}

\begin{quote}
\noindent {\small We consider Hamiltonian with $N$ point
interactions in $\R^d,\: d=2,3,$ all with the same coupling
constant, placed at vertices of an equilateral polygon $\PP_N$. It
is shown that the ground state energy is locally maximized by a
regular polygon. The question whether the maximum is global is
reduced to an interesting geometric problem.}
\end{quote}


\section{Introduction}

Questions about geometrical configurations which lead to extremal
value of a spectral quantity represent a classical topic in
mathematical physics; recall the Faber-Krahn inequality \cite{Fa,
Kr}, the PPW-conjecture proved by Ashbaugh and Benguria \cite{AB},
and numerous other examples. A particular place in this list is
occupied by the Dirichlet problem for the Laplacian in annular
strips and their higher-dimensional analogues where the principal
eigenvalue is typically maximized by a circular shape \cite{EHL}.

The reason behind the last named result is an effective attraction
coming from the curvature. This effect is robust and can be
manifested in situation where the confinement to the vicinity of a
certain geometric object is much weaker than boundary conditions,
being realized, for instance, by a potential or even by a mere
family of point interactions. An illustration is is provided by
``polymer'' models \cite{AGHH}, i.e. infinite equidistant arrays
of point interactions: if such a polymer is curved but
asymptotically straight in a suitable sense, it has a nonempty
discrete spectrum the properties of which depend substantially on
the geometry of the array \cite{Ex1, EN}.

It is natural to ask whether the mentioned results about Dirichlet
annuli have an analogue in the situation when the point
interactions are arranged along a closed curve of a fixed length.
In this paper we address this isoperimetric problem and show that
the ``circular'' shape, namely a regular polygon is a local
maximizer for the lowest eigenvalue.

On the other hand, the question about the global uniqueness of
this maximizer is left open. As we shall see in Sec.~\ref{geom}
below, the problem can be reduced to verification of a simple
property for some families of polygon diagonals. At a glance it
seems to be something which must be known since Euclid's
\emph{Elementa}, or at least for quite a long time. However, this
impression is wrong; it is found nowhere in the literature unless
I looked the wrong direction and asked wrong people. And as any
problem which allow a statement in elementary geometric terms, it
has a certain independent appeal.

We will formulate the problem and state our main result,
Theorem~\ref{main}, in the next section. It will be then proved in
the Secs~\ref{geom} and \ref{s: local}, while the last two
sections are devoted to the global uniqueness question and
possible extensions of the result.


\section{The main result}
\setcounter{equation}{0}

Let $\PP_N\subset\R^d,\: d=2,3,$ be a polygon which it is for the
present purpose convenient to identify with an ordered set of its
vertices, $\PP_N=\{y_1,\dots,y_N\}$; if the vertex indices exceed
this range they are understood $\mathrm{mod\,}N$. We suppose that
$\PP_N$ is \emph{equilateral}, $|y_{i+1}-y_i|=\ell\:$ for a fixed
$\ell>0$ and any $i$. By $\tilde\PP_N$ we denote a \emph{regular}
polygon of the edge length $\ell$, which means planar (this is
trivial if $d=2$) with vertices lying on a circle of radius
$\ell\left( 2\sin \frac{\pi}{N}\right)^{-1}$.

The object of our interest is the Hamiltonian $-\Delta_{\alpha,
\PP_N}$ in $L^2(\R^d)$ with $N$ point interactions, all of the
same coupling constant $\alpha$, placed at the vertices of
$\PP_N$. We suppose that this operator has a non-empty discrete
spectrum,
 $$
 \epsilon_1 \equiv \epsilon_1(\alpha,\PP_N):= \inf \sigma
 \left(-\Delta_{\alpha,\PP_N}\right)<0\,,
 $$
which is satisfied for any $\alpha\in\R$ if $d=2$, while in the
case $d=3$ it is true below a certain critical value of $\alpha$
-- cf.~\cite[Sec.~II.1]{AGHH}.

 \begin{theorem} \label{main}
 Under the stated conditions, $\epsilon_1(\alpha,\PP_N)$ is for
 fixed $\alpha$ and $\ell$ locally sharply maximized by a regular
 polygon, $\PP_N= \tilde\PP_N$.
 \end{theorem}

\noindent Let us remark that speaking about uniqueness of the
maximizer, we have of course in mind the family of regular
polynomials related mutually by Euclidean transformations of
$\R^d$.


\section{A geometric reformulation} \label{geom}
\setcounter{equation}{0}

As the first step to prove Theorem~\ref{main} we want to show that
the task can be reduced to a geometric problem. Using the standard
notation, $k=i\kappa$ with $\kappa>0$, we find the eigenvalues
$-\kappa^2$ solving the following spectral condition,
 $$
 \det \Gamma_k =0 \qquad \mathrm{with} \qquad (\Gamma_k)_{ij}:=
 (\alpha-\xi^k)\delta_{ij} - (1-\delta_{ij}) g^k_{ij}\,,
 $$
where $g^k_{ij}:= G_k(y_i-y_j)$, or equivalently
 \begin{equation} \label{g}
 g^k_{ij} = \left\{ \begin{array}{ccc} \frac{1}{2\pi}
 K_0(\kappa|y_i-y_j|) &\quad\dots\quad& d=2 \\ [.3em]
 \frac{e^{-\kappa|y_i-y_j|}}{4\pi|y_i-y_j|}
 &\quad\dots\quad& d=3 \end{array} \right.
 \end{equation}
and the regularized Green's function at the interaction site is
 $$
 \xi^k = \left\{ \begin{array}{ccc} -\frac{1}{2\pi}
 \left(\ln\frac{\kappa}{2} +\gamma_\mathrm{E} \right)
 &\quad\dots\quad& d=2 \\ [.3em]
 -\frac{\kappa}{4\pi} &\quad\dots\quad& d=3 \end{array} \right.
 $$
The matrix $\Gamma_{i\kappa}$ has $N$ eigenvalues counting
multiplicity which are decreasing in $(-\infty,0)$ as functions of
$\kappa$ -- see \cite{KL} and recall that they are real-analytic
and non-constant in view of their known asymptotic behavior
\cite{AGHH}. The quantity in question, $\epsilon_1(\alpha,\PP_N)$,
corresponds to the point $\kappa$ where the lowest of these
eigenvalues vanishes. Consequently, we have to check that
 \begin{equation} \label{Gammaineq}
 \min \sigma(\Gamma_{i\tilde\kappa_1}) < \min
 \sigma(\tilde\Gamma_{i\tilde\kappa_1})
 \end{equation}
holds locally for $\PP_N\ne \tilde \PP_N$, where
$-\tilde\kappa_1^2 = \epsilon_1(\alpha,\tilde\PP_N)$.

Next we notice that the lowest eigenvalue of
$\tilde\Gamma_{i\tilde\kappa_1}$ corresponds to the eigenvector
$\tilde\phi_1= N^{-1/2}(1,\dots,1)$. Indeed, by \cite{AGHH} there
is a bijective correspondence between an eigenfunction $c=(c_1,
\dots,c_N)$ of $\Gamma_{i\kappa}$ at the point, where the
corresponding eigenvalue equals zero, and the corresponding
eigenfunction of $-\Delta_{\alpha,\PP_N}$ given by $c
\leftrightarrow \sum_{j=1}^N c_j G_{i\kappa}(\cdot-y_j)$, up to a
normalization. Again by \cite{AGHH}, the principal eigenvalue of
$-\Delta_{\alpha,\PP_N}$ is simple, so it has to be associated
with a one-dimensional representation of the corresponding
discrete symmetry group of $\tilde\PP_N$; it follows that
$c_1=\dots=c_N$. Hence
 \begin{equation} \label{minGammatilde}
 \min \sigma(\tilde\Gamma_{i\tilde\kappa_1}) = (\tilde\phi_1,
 \tilde\Gamma_{i\tilde\kappa_1} \tilde\phi_1) = \alpha -
 \xi^{i\tilde\kappa_1} - \frac{2}{N} \sum_{i<j}
 \tilde g_{ij}^{i\tilde\kappa_1}\,.
 \end{equation}
On the other hand, for the l.h.s. of (\ref{Gammaineq}) we have a
variational estimate
 $$
 \min \sigma(\Gamma_{i\tilde\kappa_1}) \le (\tilde\phi_1,
 \Gamma_{i\tilde\kappa_1} \tilde\phi_1) = \alpha -
 \xi^{i\tilde\kappa_1} - \frac{2}{N} \sum_{i<j}
 g_{ij}^{i\tilde\kappa_1}\,,
 $$
and therefore it is sufficient to check that the inequality
 \begin{equation} \label{Greenineq}
 \sum_{i<j} G_{i\kappa}(y_i-y_j) > \sum_{i<j}
 G_{i\kappa}(\tilde y_i-\tilde y_j)
 \end{equation}
holds \emph{for all} $\kappa>0$ and $\PP_N\ne \tilde \PP_N$ in the
vicinity of the regular polygon $\tilde \PP_N$.

For brevity we introduce the symbol $\ell_{ij}$ for the diagonal
length $|y_i-y_j|$ and $\tilde\ell_{ij}:=|\tilde y_i- \tilde
y_j|$. We define the function $F:\: (\R_+)^{N(N-3)/2} \to\R$ by
 $$
 F(\{\ell_{ij}\}): = \sum_{m=2}^{[N/2]}\: \sum_{|i-j|=m}
 \left[ G_{i\kappa}(\ell_{ij}) -G_{i\kappa}(\tilde\ell_{ij})
 \right]\,;
 $$
notice that $m=1$ does not contribute due to the assumed
equilaterality of $\PP_N$. Our aim is to show that
$F(\{\ell_{ij}\})>0$ except if $\{\ell_{ij}\}=
\{\tilde\ell_{ij}\}$. We use the fact that the function
$G_{i\kappa}(\cdot)$ is \emph{convex} for any fixed $\kappa>0$ and
$d=2,3$ as it can be seen from cf.~(\ref{g}); this yields the
inequality
 $$
 F(\{\ell_{ij}\})\ge  \sum_{m=2}^{[N/2]} \nu_m
 \left[ G_{i\kappa}\left(\frac{1}{\nu_m} \sum_{|i-j|=m}
 \ell_{ij} \right) -G_{i\kappa}(\tilde\ell_{1,1+m}) \right]\,,
 $$
where $\nu_n$ is the number of the appropriate diagonals,
 $$
 \nu_m:= \left\{ \begin{array}{ccl} N &\quad\dots\quad&
 m=1,\dots, \left[\frac{1}{2}(N-1)\right] \\ [.3em]
 \frac{1}{2}N &\quad\dots\quad& m=\frac{1}{2}N
 \quad\; \mathrm{for}\; N \;\mathrm{even} \end{array} \right.
 $$
At the same time, $G_{i\kappa}(\cdot)$ is monotonously decreasing
in $(0,\infty)$, so the sought claim would follow if we
demonstrate the inequality
 $$
 \tilde\ell_{1,m+1} \ge \frac{1}{\nu_n} \sum_{|i-j|=m}
 \ell_{ij}
 $$
and show that it is sharp for at least one value of $m$ if
$\PP_N\ne \tilde \PP_N$.

Thus we have managed to reformulate our problem in purely
geometric terms. Since the corresponding property -- to be checked
in the following -- may be of independent interest we will state
it more generally, without dimensional restrictions. Let $\PP_N$
be an equilateral polygon in $\R^d,\, d\ge 2$. Given a fixed
integer $m=2,\dots, [\frac{1}{2}N]$ we denote by $\DD_m$ the
\emph{sum of lengths of all $m$-diagonals}, i.e. the diagonals
jumping over $m$ vertices.
 \begin{description}
 \item\emph{($P_m$)} The quantity $\DD_m$ is, in the set of
 equilateral polygons $\PP_N\subset\R^d$ with a fixed edge length
 $\ell>0$, uniquely maximized by $\tilde \DD_m$ referring to the
 (family of) regular polygon(s) $\tilde\PP_N$.
 \end{description}


\section{A local maximizer} \label{s: local}
\setcounter{equation}{0}

Our next goal is to demonstrate the following claim which yields
in the particular cases $d=2,3$ our main result,
Theorem~\ref{main}.

 \begin{theorem} \label{local}
 The property ($P_m$) holds locally for any $m=2,\dots,
 [\frac{1}{2}N]$.
 \end{theorem}

\smallskip

\noindent
\begin{proof} We have to find, for instance, local maxima
of the function
 $$
 f_m:\: f_m(y_1,\dots,y_N) = \frac{1}{N}\sum_{i=1}^N |y_i-y_{i+m}|
 $$
under the constraints $g_i(y_1,\dots,y_n)=0$, where
 $$
 g_i(y_1,\dots,y_n):= \ell- |y_i-y_{i+1}|\,, \quad i=1,\dots,N\,.
 $$
The number of independent variables is in fact $(N-2)(d-1)-1$
because $2d-1$ parameters are related to Euclidean transformations
and can be fixed. We put
 \begin{equation} \label{Km}
 K_m(y_1,\dots,y_N) := f_m(y_1,\dots,y_N) + \sum_{r=1}^N
 \lambda_r g_r(y_1,\dots,y_n)
 \end{equation}
and compute the derivatives $\nabla_j K_m(y_1,\dots,y_N)$ which
are equal to
 $$
 \frac{1}{N} \left\{
 \frac{y_j-y_{j+m}}{|y_j-y_{j+m}|}
 + \frac{y_j-y_{j-m}}{|y_j-y_{j-m}|} \right\}
 - \lambda_j \frac{y_j-y_{j+1}}{\ell}
 - \lambda_{j-1} \frac{y_j-y_{j-1}}{\ell}\,.
 $$
We want to show that these expressions vanish for a regular
polygon. Let us introduce a parametrization for any planar
equilateral polygon. Without loss of generality we may suppose
that it lies in the plane of the first two axes. The other
coordinates are then zero and we neglect them writing
 \begin{equation} \label{planpoly}
 y_j= \ell \left( \sum_{n=0}^{j-1} \cos\left( \sum_{i=1}^n \beta_i -
 \varphi \right), \sum_{n=0}^{j-1} \sin\left( \sum_{i=1}^n \beta_i -
 \varphi \right) \right)\,,
 \end{equation}
where $\varphi\in\R$ is a free parameter and $\beta_i$ is the
``bending angle'' at the $i$th vertex (modulo $2\pi$); the family
of these angles satisfies naturally the condition
 \begin{equation} \label{anglenorm}
 \sum_{i=1}^N \beta_i = 2\pi w
 \end{equation}
for some $w\in\Z$. Choosing $\tilde\varphi= \frac{\pi}{N}$ and
$\tilde\beta_i= \frac{2\pi i}{N}$, we get in particular
 $$
 \tilde y_{\pm m}= \ell \left( \pm\sum_{n=0}^{m-1}
 \cos\frac{\pi}{N}(2n+1), \sum_{n=0}^{m-1}
 \sin\frac{\pi}{N}(2n+1) \right)\,.
 $$
Then we have
 $$
 |\tilde y_j-\tilde y_{j\pm m}|= \ell \left[\left( \sum_{n=0}^{m-1}
 \cos\frac{\pi}{N}(2n+1) \right)^2 + \left(\sum_{n=0}^{m-1}
 \sin\frac{\pi}{N}(2n+1) \right)^2 \right] =: \ell \Upsilon_m\,,
 $$
and consequently, $\nabla_j K_m(\tilde y_1,\dots,\tilde y_N)=0$
holds for $j=1,\dots,N$ if we choose all the Lagrange multipliers
in (\ref{Km}) equal to
 \begin{equation} \label{lagrange}
 \lambda = \frac{\sigma_m}{N\Upsilon_m} \qquad \mathrm{with}\qquad
 \sigma_m := \frac{\sum_{n=0}^{m-1} \sin\frac{\pi}{N}(2n+1)}
 {\sin\frac{\pi}{N}} = \frac{\sin^2\frac{\pi m}{N}}
 {\sin^2\frac{\pi}{N}}\,.
 \end{equation}
The second partial derivatives, $\nabla_{k,r}\nabla_{j,s}
K_m(y_1,\dots,y_N)$, are computed to be
 \begin{eqnarray*} 
 \lefteqn{ \frac{1}{N} \Bigg\{ \frac{\delta_{kj}- \delta_{k,j+m}}
 {|y_j-y_{j+m}|} \delta_{rs} - \frac{(y_j-y_{j+m})_r
 (y_j-y_{j+m})_s (\delta_{kj}- \delta_{k,j+m})} {|y_j-y_{j+m}|^3}
 + \frac{\delta_{kj}- \delta_{k,j-m}}
 {|y_j-y_{j-m}|} \delta_{rs} } \\ &&  - \frac{(y_j-y_{j-m})_r
 (y_j-y_{j-m})_s (\delta_{kj}- \delta_{k,j-m})} {|y_j-y_{j-m}|^3}
 + \frac{\lambda}{\ell} \Big(\delta_{k,j+m}+ \delta_{k,j-m}
 -2\delta_{kj}\Big) \delta_{rs}\Bigg\}\,. \phantom{AA}
 \end{eqnarray*}
This allows us to evaluate the Hessian at the stationary point.
After a long but straightforward calculation we arrive at the
expression
 \begin{eqnarray} \label{hess}
 \lefteqn{ \sum_{k,j,r,s} \nabla_{k,r}\nabla_{j,s}
 K_m(\tilde y_1,\dots,\tilde y_N) \xi_{k,r} \xi_{j,s}} \\ &&
 = \frac{1}{N\ell\Upsilon_m} \sum_{j=1}^N
 \left\{ |\xi_j-\xi_{j+m}|^2 -\frac{(\xi_j-\xi_{j+m},
 \tilde y_j-\tilde y_{j+m})^2}{|\tilde y_j-\tilde y_{j+m}|^2}
 - \sigma_m |\xi_j-\xi_{j+1}|^2 \right\}\,. \nonumber \phantom{AA}
 \end{eqnarray}
We observe that the form depends on vector differences only, so it
is invariant with respect to Euclidean transformations.
Furthermore, the sum of the first two terms in the bracket at the
r.h.s. of (\ref{hess}) is non-negative by Schwarz inequality.

Since the second term in non-positive, it will be sufficient to
establish negative definiteness of the quadratic form
 \begin{equation} \label{form}
 \xi\mapsto S_m[\xi]:= \sum_j \left\{ |\xi_j-\xi_{j+m}|^2
 - \sigma_m |\xi_j-\xi_{j+1}|^2 \right\}
 \end{equation}
on $\R^{Nd}$. Moreover, it is enough to consider here the case
$d=1$ only because $S_m$ is a sum of its ``component'' forms. We
observe that the matrices corresponding to the two parts of
(\ref{form}) can be simultaneously diagonalized; the corresponding
eigenfunctions are $\{{\sin\choose\cos} (\mu_r j)\}_{j=1}^N$,
where $\mu_r= \frac{2\pi r}{N},\; r=0,1,\dots,m-1$. Taking the
corresponding eigenvalues we see that it is necessary to establish
the inequalities
 \begin{equation} \label{ev-ineq}
 4\left\{ \sin^2 \frac{\pi mr}{N} -\sigma_m \sin^2 \frac{\pi r}{N}
 \right\} < 0
 \end{equation}
for $m=2,\dots,[\frac{1}{2}N]$ and $r=2,\dots,m-1$. We left out
here the case $r=1$ when the l.h.s. of (\ref{ev-ineq}) vanishes.
At the same time, however, the above explicit form of the
eigenfunctions shows that the corresponding $\xi_j-\xi_{j+m}$ are
in this case proportional to $\tilde y_j-\tilde y_{j+m}$ so the
second term at the r.h.s. of (\ref{hess}) is negative unless
$\xi=0$.

Using the expression (\ref{lagrange}) for $\sigma_m$ we can
rewrite the condition (\ref{ev-ineq}) in terms of Chebyshev
polynomials of the second kind as
 \begin{equation} \label{ev-ineq2}
 U_{m-1}\left(\cos\frac{\pi}{N}\right) > \left|
 U_{m-1}\left(\cos\frac{\pi r}{N}\right)\right|\,,
 \end{equation}
which can be checked using properties of these polynomials
\cite[Chap.~22]{AS}. One can do it also directly, because
(\ref{ev-ineq2}) is equivalent to
 $$
 \sin\frac{\pi m}{N} \sin\frac{\pi r}{N} > \left|
 \sin\frac{\pi}{N} \sin\frac{\pi mr}{N} \right|\,, \qquad
 2\le r<m \le \left[\frac N2 \right]\,.
 $$
We have $\sin x\: \sin(\eta^2/x)\ge \sin\eta$ for a fixed $\eta\in
(0,\frac 12\pi)$ and $2\eta^2/\pi \le x \le \frac 12\pi$, and
moreover, this inequality is sharp if $x\ne\eta$, hence the
desired assertion follows from the inequality $\sin^2 x -
\sin\frac{\pi}{N} \sin\frac{Nx^2}{\pi} \ge 0$ valid for $x\in
(0,\frac 12\pi)$. This concludes the proof of Theorem~\ref{local},
and by that also of Theorem~\ref{main}.
\end{proof}


\section{Global properties}
\setcounter{equation}{0}

The question whether the maximizer represented by regular polygons
is global at the same time is more difficult. By the argument of
Sec.~\ref{geom} it can be reduced again to a purely geometric
problem, namely that about validity of the following claim.

 \begin{conjecture} \label{global}
 The property ($P_m$) holds globally for any $m=2,\dots,
 [\frac{1}{2}N]$.
 \end{conjecture}

\noindent Let us look at the problem in more detail in the
particular case of \emph{planar polygons}, $d=2$. We employ a
parametrization analogous to (\ref{planpoly}): for a fixed $i$ we
identify $y_i$ with the origin and set for simplicity $\varphi=0$,
i.e.
 $$
 y_{i+m}= \ell \left( 1+ \sum_{n=1}^{m-1} \cos \sum_{j=1}^n \beta_{j+i},\:
 \sum_{n=1}^{m-1} \sin \sum_{j=1}^n \beta_{j+i} \right)\,;
 $$
in addition to the angular condition (\ref{anglenorm}) we require
naturally also that $y_i=y_{i+N}$, or in other words
 \begin{equation} \label{winding}
 1+ \sum_{n=1}^{N-1} \cos \sum_{j=1}^n \beta_{j+i} =
 \sum_{n=1}^{N-1} \sin \sum_{j=1}^n \beta_{j+i} = 0
 \end{equation}
for any $i=1,\dots,N$. The mean length of all $m$-diagonals is
easily found,
 $$
 M_m= \frac{\ell}{N}\sum_{i=1}^N \left[ \left( 1+ \sum_{n=1}^{m-1}
 \cos \sum_{j=1}^n \beta_{j+i} \right)^2 + \left( \sum_{n=1}^{m-1}
 \sin \sum_{j=1}^n \beta_{j+i} \right)^2 \right]^{1/2},
 $$
or alternatively
 \begin{equation} \label{meandiag2}
 M_m= \frac{\ell}{N}\sum_{i=1}^N \left[ m+ 2\sum_{n=1}^{m-1}
 \sum_{r=1}^n \cos \sum_{j=r}^n \beta_{j+i} \right]^{1/2}.
 \end{equation}
It allows us to prove the claim in the simplest nontrivial case.

 \begin{proposition} \label{m=2}
 The property ($P_2$) holds globally if $\:d=2$.
 \end{proposition}

\smallskip

\noindent
\begin{proof}
By (\ref{meandiag2}) the mean length of the $2$-diagonals equals
 $$
 M_2= \frac{\sqrt{2}\ell}{N}\sum_{i=1}^N (1+\cos\beta_i)^{1/2} =
 \frac{2\ell}{N}\sum_{i=1}^N \cos\frac{\beta_i}{2}\,;
 $$
notice that $\cos\frac{\beta_i}{2}>0$ because $\beta_i\in
(-\pi,\pi)$. Using now convexity of the function $u\mapsto
-\cos\frac{u}{2}\:$ in $(-\pi,\pi)$ together with the condition
(\ref{anglenorm}) we find
 $$
 - \sum_{i=1}^N \cos\frac{\beta_i}{2} \ge -N \cos\left(
 \sum_{i=1}^N \frac{\beta_i}{2} \right) = - N \cos\frac{\pi}{N}\,,
 $$
and therefore $M_2 \le 2\ell \cos\frac{\pi}{N}= \tilde M_2$.
Moreover, since the said function is strictly convex, the
inequality is sharp unless all the $\beta_i$'s are the same.
\end{proof}

For $m\ge 3\:$ the situation is more complicated and one has to
take into account also the condition (\ref{winding}); for the
moment the problems remains open.


\section{Possible extensions}
\setcounter{equation}{0}

Apart of proving Conjecture~\ref{global} and by that the global
uniqueness of the maximizer, the present problem offers various
other extensions. One can ask, for instance, what will be the
maximizer when we replace the equilaterality by a prescribed
ordered $N$-tuple of polygon lengths $\{\ell_j\}$ and/or coupling
constants $\{\alpha_j\}$. In both cases the task becomes more
difficult because we loose the ground state symmetry which yielded
the relation (\ref{minGammatilde}) and consequently the geometric
reformulation based on the inequality (\ref{Greenineq}).

One can also attempt to extend the result to point interaction
family of point interactions in $\R^3$ placed on a closed surface.
In this case, however, there is no unique counterpart to the
equilaterality and one has to decide first what the ``basic cell''
of such a polyhedron surface should be. Another extensions of our
isoperimetric problem concern ``continuous'' versions of the
present situation, i.e. Schr\"odinger operators with singular
interactions supported by closed curves or surfaces --
cf.~\cite{EI, Ex2} and references therein -- or with a regular
potential well extended along a closed curve.


\subsection*{Acknowledgments}

The research has been partially supported by ASCR within the
project K1010104.

\end{document}